\documentclass[pre,aps,showpacs,twocolumn]{revtex4}
\usepackage{graphicx}

\begin{document}
\newcommand{\beq}{\begin{equation}}
\newcommand{\eeq}{\end{equation}}
\newcommand{\barr}{\begin{eqnarray}}
\newcommand{\earr}{\end{eqnarray}}

\newcommand{\andy}[1]{}



\def\cE{{\mathcal{E}}}


\title{Fractal entropy of a chain of nonlinear oscillators}
\author{A. Scardicchio}
\altaffiliation[Present address: ]{Center for Theoretical Physics,
Massachusetts Institute of Technology, Cambridge, MA 02139.}
\affiliation{Dipartimento di Fisica, Universit\`a di Bari, I-70126
Bari, Italy}
\affiliation{Istituto Nazionale di Fisica Nucleare, Sezione di
Bari, I-70126  Bari, Italy}

\author{P. Facchi}
\affiliation{Dipartimento di Fisica, Universit\`a di Bari, I-70126
Bari, Italy}
\affiliation{Istituto Nazionale di Fisica Nucleare, Sezione di
Bari, I-70126  Bari, Italy}
\author{S. Pascazio} 
\affiliation{Dipartimento di Fisica, Universit\`a di Bari, I-70126
Bari, Italy}
\affiliation{Istituto Nazionale di Fisica Nucleare, Sezione di
Bari, I-70126  Bari, Italy}

\date{\today}

\begin{abstract}
We study the time evolution of a chain of nonlinear oscillators.
We focus on the fractal features of the spectral entropy and
analyze its characteristic intermediate timescales as a function
of the nonlinear coupling. A Brownian motion is recognized, with
an analytic power-law dependence of its diffusion coefficient on
the coupling.
\end{abstract}


\maketitle

\section{\label{sec:introduction} Introduction}
A system composed of a large number of particles (generically)
relaxes toward an equilibrium state that is independent of the
details of the initial state. This is one of the fundamental
hypotheses of statistical mechanics. However, from the point of
view of Hamiltonian dynamics, the detailed features of the
relaxation are not thoroughly understood. One of the unsolved
fundamental questions is how equilibrium is approached when the
underlying microscopic dynamics is sufficiently chaotic, without
introducing any randomization or coarsegraining ``by hand."

After the pioneering work on the time evolution of nonintegrable
systems \cite{FPU} and ergodicity \cite{Sinai}, it is now clear
that several additional factors play a primary role in
characterizing the dynamical evolution
\cite{KAM,Nekhoroshev,Arnolddiff,Lichtenberg}.
However, for a large number of particles the KAM argument becomes
less effective
\cite{Lichtenberg,ParisiSFT} and the Nekhoroshev bound \cite{Nekhoroshev}
for the equilibrium time appears too weak in comparison with
numerical results. This scenario has motivated a number of
numerical studies of Fermi-Pasta-Ulam-like models, in the attempt
to clarify the dependence of the equilibrium time (defined in
terms of suitable indicators) on the strength of the nonlinear
terms and the number of particles
\cite{Casetti97}. These studies yield
stretched-exponential relaxation laws, enforcing the picture that
macroscopic equilibrium could be built out of local ones
\cite{Reigada02}.

The aim of this article is to investigate this issue by analyzing,
both analytically and numerically, the dynamics of a chain of $N$
coupled anharmonic oscillators at intermediate timescales (for
states that are close to equilibrium). One of our main results is
that at these intermediate timescales the system performs a
Brownian motion with a diffusion constant that can be accurately
estimated and turns out to be {\em analytically} diverging in the
coupling constant: as a consequence, a perturbative approach to
this problem appears sensible.

\section{\label{sec:thesystem} The system}
We will study a Hamiltonian made up of an integrable part and a
(small) nonlinear perturbation. This is a classical nonintegrable
system, that does not possess enough integrals of motion. The
conjugate variables are $(q,p)=(q_1,...q_N,p_1,...,p_N)$, where
periodic boundary conditions $q_{N+1}=q_1,\; p_{N+1}=p_1$ are
understood. We take $N=2^7=128$, for convenience of the numerical
algorithm (we observed no significant difference for larger $N$).
The Hamiltonian ($\phi^4$ model) reads
\andy{eq:hamilttot,eq:hamiltf}
\barr
\label{eq:hamilttot}
&& H(q,p)=H_0(q,p)+g V(q), \quad (g>0)
\\
\label{eq:hamiltf}
& & H_0(q,p)=\sum_{i=1}^N \frac{1}{2}p_i^2+\frac{1}{2}m^2
q_i^2+\frac{1}{2}\left(q_{i+1}-q_i\right)^2,
\\
& &  V(q)=\sum_{i=1}^N
\frac{1}{4}q_i^4.
\earr
The quadratic part $H_0$ is easily diagonalized by means of a
discrete Fourier transform, in terms of the $2N$ normal variables
$q_\kappa,\ p_\kappa$, with $\kappa=(k,a)$, where $k=0,...,N/2$
and $a=0,1$, $a=0\; (a=1)$ corresponding to the cosine (sine)
transform with wave number $k$. With this coordinate change $H_0$
becomes
\andy{eq:H2}
\beq
H_0=\sum_\kappa E_\kappa,\qquad E_\kappa = \frac{1}{2}p_\kappa^2 +
\frac{1}{2}\omega_k^2 q_\kappa^2,
\label{eq:H2}
\eeq
with the frequency spectrum
\andy{eq:Hspec}
\barr
\omega_k^2 &=&m^2+2\left(1-\cos\left(\frac{2\pi k}{N}\right)\right),
\nonumber \\
m &=& \omega_{\mathrm{min}} \leq \omega_k\leq
\omega_{\mathrm{MAX}} = \sqrt{4+m^2}.
\label{eq:Hspec}
\earr
In this article we always set $m^2 \gtrsim 0.1$. The value of $m$
determines the width of the spectrum and has profound consequences
on the dynamics, in particular at small nonlinearities: metastable
states, like solitons and breathers, are born more easily at small
$m$ and this can have drastic consequences, both at intermediate
and large timescales of the equilibration process
\cite{Reigada02}.

We integrated the Hamilton equations deriving from
(\ref{eq:hamilttot}) via a fourth-order Runge-Kutta algorithm in
double precision. Energy conservation is verified at least up to 1
part in $10^7$ during the whole running time. Such a precision is
necessary in order to assure that the fluctuations due to
numerical integrations be negligible with respect to the physical
ones in which we are interested.

\section{\label{sec:specentro} Spectral entropy}
{}From the numerically integrated solutions we analyze the
behavior of the spectral entropy
\cite{Casetti97}
\andy{def:spectrentro}
\beq
\label{def:spectrentro}
S(t)=-\sum_\kappa \frac{E_\kappa}{E_0}\ln\frac{E_\kappa}{E_0},
\eeq
where $E_0=H_0$ is the unperturbed total energy. When $g=0$ all
the $E_\kappa$s, and therefore $S$, are constant. As soon as the
nonlinearity is switched on, $g>0$, the spectral entropy becomes a
nontrivial function of time.

The purpose of this letter is to explore the dynamics of the
system over \emph{intermediate} timescales. Previous studies
\cite{ParisiEPL,Casetti97} mainly concentrated on the
long-time behavior of the equilibration process, starting from
states that are far to equilibrium and smoothing $S$ over
sufficiently long time intervals. Relaxation from states that are
close to equilibrium has been more seldom studied \cite{Lepri}. In
our simulation we always set the initial conditions with the
actions (unperturbed energies) randomly picked from a
microcanonical ensemble and the angles completely random.
Therefore, the fluctuations of $S(t)$ will be studied \emph{close
to equilibrium}.

$S$ displays wild time fluctuations, over a wide range of
frequencies. We will show that useful, univocal information can be
obtained from such an irregular function: we will first recognize
a Brownian structure and then look at the characteristic
timescales and study their dependence on the nonlinear coupling.

Let us first discuss some analytical properties. At equilibrium,
average quantities and statistical properties should depend only
on integrals of motion. For the Hamiltonian (\ref{eq:hamilttot})
one argues that the only global integral of motion is the total
energy $E=H$ or, equivalently, the energy per mode $\epsilon\equiv
E/N$.  The equations for $q_i$ possess the following scaling
symmetry: if $q_i$ is a solution of the Hamilton equations with
coupling $g$, then $q'_i=\sqrt{A}q_i$ is solution of the Hamilton
equations with coupling $g'=g/A$. With this rescaling, the energy
is changed to $E'=AE$. A function $X$, representing the average of
some quantity at equilibrium and having dimension
$[X]=\text{length}^\nu$, can only depend on $E$ and $g$, whence
\andy{eq:scale}
\beq
\label{eq:scale}
X(AE,g/A)=A^{\nu/2}X(E,g).
\eeq
Therefore, if $\nu=0$, $X$ depends only on the dimensionless
product
\andy{eq:xdef}
\beq
\label{eq:xdef}
x=g\epsilon=g E/N,
\eeq
that may be considered as the effective strength of the
nonlinearity. We will study the dynamical properties of our system
for $10^{-4} \leq x \leq 1$.

Average quantities in the weakly nonlinear regime can be
calculated by using the microcanonical distribution. The use of
this distribution is motivated by the empirical observation that
the total \emph{unperturbed} energy $E_0$, although not constant
in time, fluctuates less than $1\%$ around its mean value (even
smaller fluctuations are observed for very small nonlinearities).
We can conclude that the primary role of the perturbation $V$ is
only to allow ``collisions" between normal modes (phonon
exchange), provoking in this way the transition to equilibrium
\footnote{In action-angle variables one can renormalize the free
Hamiltonian with a positive, nonlinear correction of order $x$.
The total Hamiltonian is accordingly split into a new free part
and a new perturbation.}, without ``storing" any significant
energy. Therefore we can assume $E\simeq E_0$ in the following.

For the spectral entropy at microcanonical equilibrium one finds,
after a straightforward but somewhat lengthy calculation,
\andy{eq:meanS,microcandeltas2}
\beq
\label{eq:meanS}
\overline{S}=\psi(N+1)-(1-\gamma)=\ln
N-(1-\gamma)+O\left(\frac{1}{N}\right),
\eeq
\barr
\label{eq:microcandeltas2}
\delta S^2&\equiv&
\overline{S^2}-\overline{S}^2=\frac{3+3N+(\pi^2-6)N^2}{3N^2(N+1)}-\psi'(N)\nonumber\\
&=& \frac{0.289}{N}+O\left(\frac{1}{N^2}\right),
\earr
where $\psi$ is the Euler digamma function \cite{Abramovitz} and
$\gamma\simeq0.5772$ the Euler-Mascheroni constant. The asymptotic
behavior of (\ref{eq:meanS}) is in agreement with previous results
\cite{Licht1,Lichtenberg}, obtained at canonical equilibrium.
In fact, as shown in the Appendix, the microcanonical and
canonical averages of any function of the spectral entropy
coincide. Equations (\ref{eq:meanS}) and
(\ref{eq:microcandeltas2}) are therefore valid also at canonical
equilibrium. These analytical results are well confirmed by
numerical simulations and provide a good test of the fact that our
numerical sample was representative of an equilibrium situation
and free from ``trend" components.

\section{\label{sec:corrfunc} Correlation function}
We study the fractal dimension and the characteristic timescales
of the entropy by looking at the correlation function $C$ for the
generalized Brownian process $S$:
\andy{eqdefC}
\beq
\label{eq:eqdefC}
C(\tau)=\lim_{T\to\infty}\frac{1}{T}\int_0^T dt\;
\left(S(t+\tau)-S(t)\right)^2.
\eeq
In general, one can identify a fractal, or generalized Brownian
motion, by the dependence $C\propto\tau^{2H}$. The exponent $H$ is
related to the fractal dimension by $D_f=2-H$. When using the
function (\ref{eq:eqdefC}) one should take care of ``detrending"
$S$ \cite{Mandelbrot}. However, since the system starts at
equilibrium, where $S$ fluctuates about its constant mean value
(\ref{eq:meanS}), no detrending is required. We emphasize,
however, that we obtained the same results also in nonequilibrium
situations (not too far from equilibrium), provided the trend
component of $S(t)$ was suitably removed.

\begin{figure}
\includegraphics[width=7.7cm]{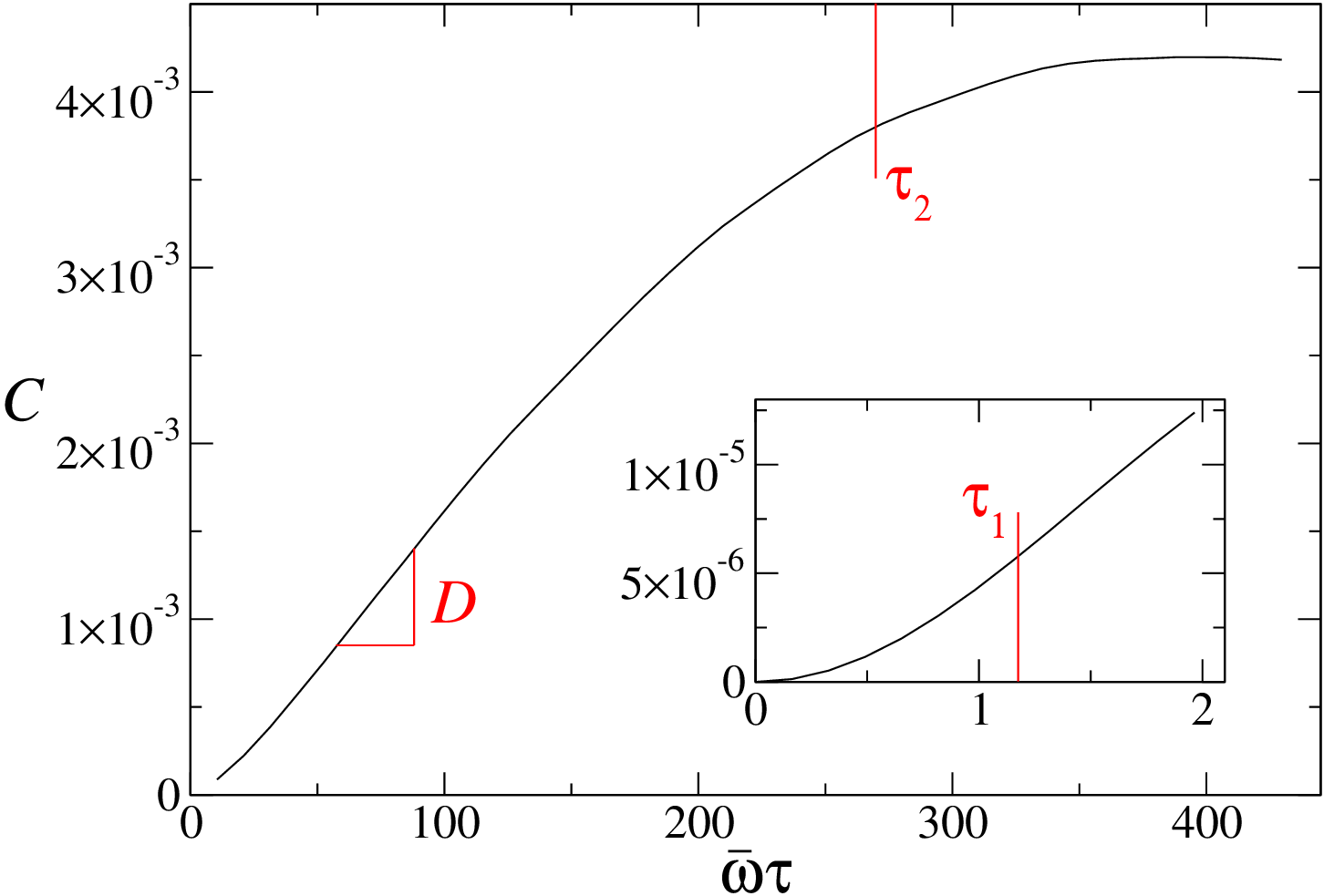}\\
 \caption{
\label{fig:corr1}
$C$ versus $\bar \omega \tau$. Notice the quadratic region
extending up to times of order $\tau_1$ (inset) and the linear
region (\ref{eq:ctaud}) in the range $\tau_1<\tau<\tau_2$. The
saturation level $2\delta S^2\simeq 0.0044$ is in full agreement
with (\ref{eq:microcandeltas2}) [for $N=128$]. We set $x=0.09$ and
$m^2=5$, so that $\bar \omega=2.62$. Observe that $\tau_2\simeq
200\; \tau_1$ is clearly an intermediate timescale, orders of
magnitude smaller than $T_{\rm relax}$. Larger ratios
$\tau_2/\tau_1$ are observed for smaller values of $x$, but the
chain of inequalities (\ref{eq:allscales}) always remains valid.}
\end{figure}

For a Brownian process one expects a linear dependence of $C$ on
$\tau$ \cite{Gardiner,Mandelbrot}, i.e.\ $D_f=3/2$,
\andy{Cproptau}
\beq
\label{eq:Cproptau}
C(\tau)\propto \tau.
\eeq
Brownian motions are useful idealizations to describe physical
processes in a simple and coherent mathematical way. However, in
order to treat an analytic function (like $S$) as a Brownian
process, one must identify (at least) one timescale, say $\tau_1$.
This timescale is such that by ``observing" the function at
timescales $\tau\lesssim\tau_1$ one obtains a smooth function,
while by observing it with a time-resolution $\tau\gg\tau_1$, one
recognizes a Brownian process. It is possible to unambiguously
identify the timescale $\tau_1$ since $C\sim\tau^2$ for
sufficiently small $\tau$:
\andy{Ctau2}
\barr
C(\tau)&\simeq&\lim_{T\to\infty}\frac{1}{T}\int_0^T dt
\left(\dot{S}(t)\tau+\frac{1}{2}\ddot{S}(t)\tau^2+O(\tau^3)\right)^2\nonumber\\
&=&\overline{\dot{S}(t)^2}\tau^2+O(\tau^4).
\label{eq:Ctau2}
\earr
At larger $\tau$ the quadratic dependence changes into the linear
one (\ref{eq:Cproptau}): the timescale at which this change takes
place is $\tau_1$. See inset in Figure \ref{fig:corr1}. It is
important to stress that $\tau_1$ is nothing but the linear
timescale for phonons, of the order of an inverse characteristic
frequency (\ref{eq:Hspec}) of the oscillators
\beq
\label{eq:Tmicro}
\frac{2\pi}{\bar \omega} = 2\pi
\frac{2}{\omega_{\mathrm{min}}+\omega_{\mathrm{MAX}}} =
\frac{4\pi}{m+\sqrt{4+m^2}} .
\eeq
Up to this timescale, one is able to observe the microscopic
details of the motion in phase space. At larger timescales, the
motion appears very irregular and the microscopic details are
lost.

For a Brownian process which is also bounded another timescale
appears, since $S$ bounded implies $C$ bounded and the law
$C\propto\tau$ must break down at a certain time. Let us call this
timescale $\tau_2$. It is easy to show that at equilibrium, for
sufficiently large $\tau$ we have $C(\tau)\simeq 2\delta S^2$, so
we can interpret $\tau_2$ as the time at which the autocorrelation
of the entropy vanishes:
\beq
\label{eq:autoC}
\langle S(t+\tau) S(t) \rangle - \langle S(t+\tau) \rangle\langle
S(t) \rangle \simeq 0, \qquad \mbox{for} \; \tau > \tau_2.
\eeq
This means that $S(t+\tau)$ and $S(t)$ can be considered
uncorrelated random numbers chosen from a sample of mean
$\overline{S}$ and variance $\delta S^2$. In terms of motion in
phase space one can argue that the system starts at $t=0$ at (or
very close to) equilibrium and at $t\sim \tau_2$ it has explored a
sufficiently large part of the equilibrium region, such that the
microcanonical averages
(\ref{eq:meanS})-(\ref{eq:microcandeltas2}) can be used. The
theoretical predictions (\ref{eq:microcandeltas2}) [with $N=128$],
(\ref{eq:Cproptau}) and (\ref{eq:Ctau2}) are very well verified by
the numerical data shown in Figure \ref{fig:corr1}.

The relaxation of the system, when it starts from a state far to
equilibrium, takes place on a timescale  $T_{\rm relax}$ that can
be defined in terms of the effective number of excited modes
\cite{Casetti97}
\beq
n_e\equiv \exp (S):
\eeq
clearly, if the system is initially far from equilibrium,
\beq
\label{eq:freq2}
\triangle n_e(T_{\rm relax})\equiv n_e(T_{\rm relax})-n_e(0)=O(N).
\eeq
On the other hand, due to
(\ref{eq:meanS})-(\ref{eq:microcandeltas2}), for a system close to
equilibrium
\beq
\label{eq:meso}
\triangle n_e(\tau_2)= e^{S+\delta S} - e^S \simeq \delta S \; e^S
= O(\sqrt{N}).
\eeq
In this sense, $\tau_2$ is an intermediate timescale,
characterizing, as we have seen, local fluctuations in phase
space.

Finally, if one considers that only a few oscillators can exchange
energy in a time $\tau_1$, one can summarize the above discussion
by writing
\barr
\label{eq:allscales}
& & \triangle n_e(\tau_1) = O(1) \ll \triangle n_e(\tau_2)
 \nonumber \\
& & = O(\sqrt{N}) \ll \triangle n_e(T_{\rm relax}) = O(N).
\earr

\section{\label{sec:diffcoeff} Diffusion coefficient
and intermediate timescales}

The presence of the linear region (\ref{eq:Cproptau}) for the
correlation function $C$ is observed in the whole range of $x$
investigated. This enables us to define a diffusion coefficient
$D$ as the rate at which $C$ increases in its linear regime, so
that in the region $\tau_1<\tau<\tau_2$ we have
\andy{eq:ctaud}
\beq
C=D\tau+C_0,
\label{eq:ctaud}
\eeq
where $D$ and $C_0$ have in general a nontrivial dependence on
$x$. In the following we will perform a systematic study of $D$,
which is the physically most relevant quantity and characterizes
the Brownian fluctuations. Notice that $C_0$ is related to the
initial quadratic region (\ref{eq:Ctau2}) and therefore depends on
the microscopic details of the motion.

\begin{figure}
\includegraphics[width=7.4cm]{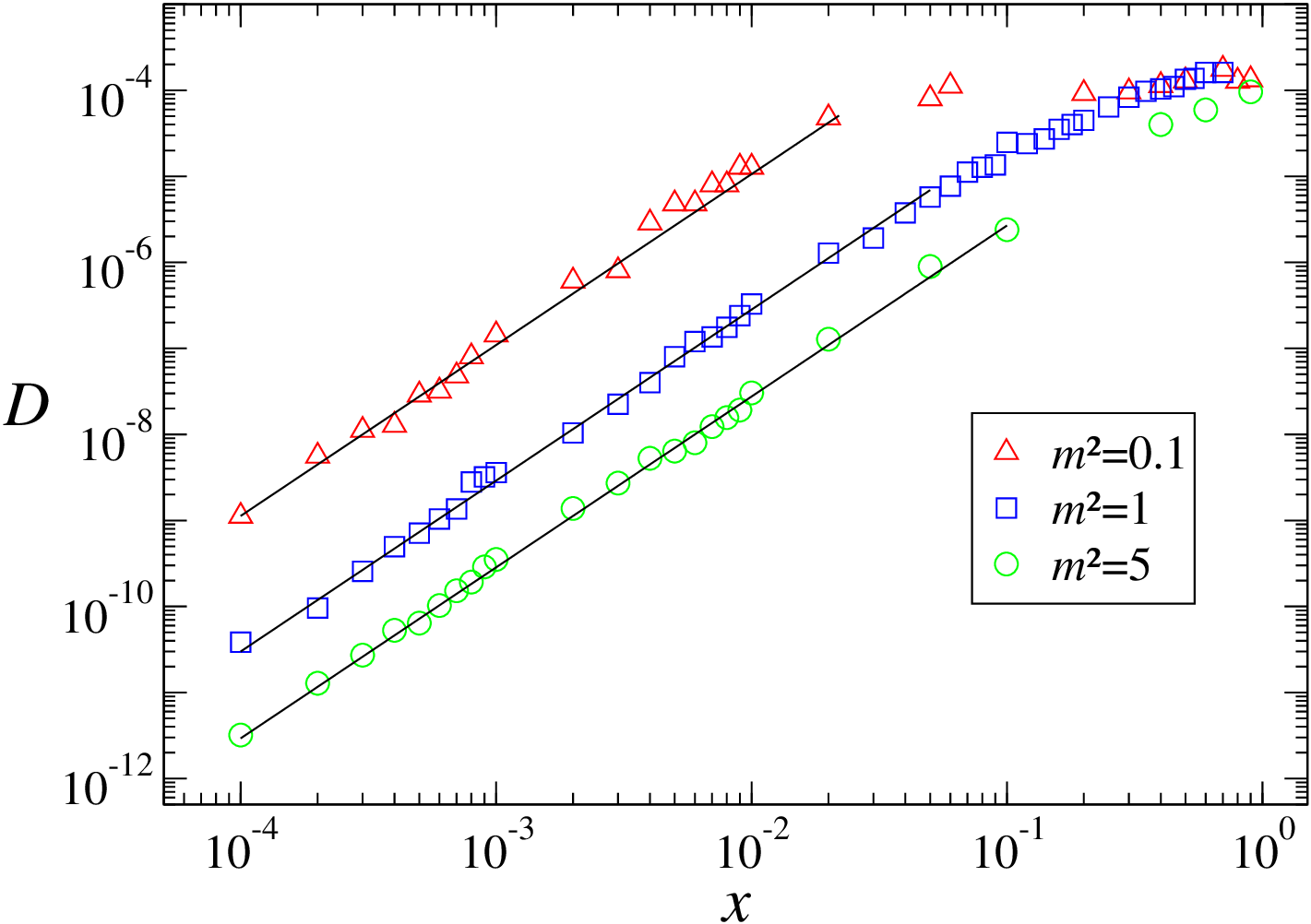}
 \caption{
\label{fig:loglogD} $D$ versus $x=g\epsilon=gE/N$ for different masses.
The lines are the power law (\ref{eq:powerlawD}) and the error
bars are included in the size of the dots. }
\end{figure}

\begin{figure}
\includegraphics[width=7.7cm]{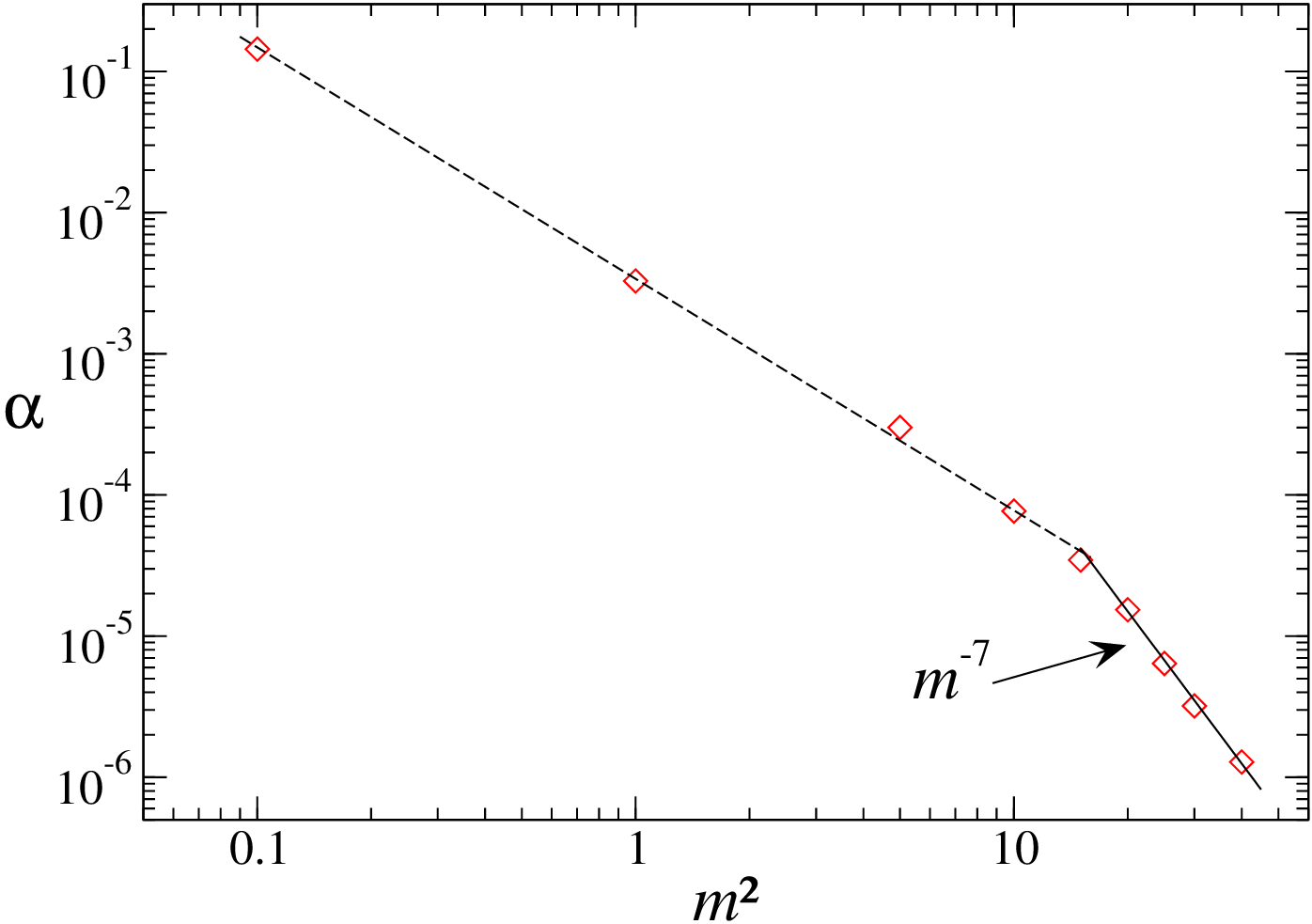}
 \caption{
\label{fig:massafit} $\alpha=D/x^2$ versus $m^2$ [in the quadratic region
(\ref{eq:powerlawD})]. The solid line is theoretical prediction
for large masses $\alpha\propto m^{-7}$. The dashed line is the
fit $\alpha\propto m^{-3.3}$. }
\end{figure}

The diffusion coefficient $D$ has no length dimension [$\nu=0$ in
(\ref{eq:scale})] and therefore one expects it to be only a
function of $x=g\epsilon=gE/N$ (at fixed $N$ and $m$). This
expectation is numerically confirmed with high accuracy. In
particular, $D$ is a monotonically increasing function of $x$, as
can be seen from Figure
\ref{fig:loglogD}. Each point in the figure is obtained by averaging
at least $5$ numerical solutions with the same value of $x$. For
$x\lesssim 0.1$, $D$ is accurately fit by a power law
\beq
D=\alpha\; x^\beta, \quad \beta =1.987\pm 0.040 .
\label{eq:powerlawD}
\eeq
This is an indication that the intermediate dynamics at these
timescales can be tackled by a perturbative approach. Notice that,
if one endeavors to fit the curves in Figure \ref{fig:loglogD}
with a stretched-exponential law of the type $D \propto \exp ({\rm
const} \cdot x^{-\delta})$, one finds the very small value $\delta
= 9.4 \cdot 10^{-4}$. In our opinion, this is  a rather strong
indication in support of a power-law behavior. On the other hand,
for large $x$, $D$ saturates to a constant ($m$-independent)
value. Observe that the coefficient $\beta$ in
(\ref{eq:powerlawD}) does {\it not} depend on $m$; the
$m$-dependence of $\alpha$ will be analyzed in the following (see
Figure \ref{fig:massafit}).

The intermediate timescale $\tau_2$ is strictly related to the
diffusion coefficient $D$, via the saturation value $2\delta S^2$
in (\ref{eq:microcandeltas2}), as discussed in connection with
Figure \ref{fig:corr1}. A consistent and natural definition is
$\tau_2=2\delta S^2/D$, so that for $x\lesssim 0.1$
\beq
\tau_2 \; \propto \;  x^{-\beta} \simeq \; x^{-2}.
\label{eq:tau2x}
\eeq
Note that in order to measure $\tau_2$ it is not even necessary to
perform a numerical integration of the Hamilton equations for
times of order $\tau_2$. It suffices to integrate them up to times
larger than $\tau_1$ ($\ll \tau_2$), get $D$ and hence $\tau_2$.

The analytic dependence (\ref{eq:tau2x}) of $\tau_2$ on $x$ is in
full agreement with previous results and suggests the validity of
a perturbative approach \cite{Lepri}. One expects that for
sufficiently small $x$ a sensible fraction of the phase space is
covered with KAM tori, allowing only Arnol'd diffusion and
yielding a consistent suppression of diffusion and a rapid
divergence of the macroscopic equilibrium time. Related analytical
and numerical work hints at a non-analytic divergence (such as a
stretched exponential) of the macroscopic timescales (such as
$T_{\rm relax}$) \cite{Nekhoroshev,ParisiSFT,Casetti97} for $x\to
0$. The fact that the intermediate timescale here analyzed
diverges with a power law (\ref{eq:tau2x}) indicates the
\emph{local} presence of a Brownian motion (in the region of
parameters studied), so that diffusion should be suppressed on
macroscopic regions of phase space.

The dependence of $\alpha=D/x^2$ on $m$ is not trivial. However,
in the region of validity of (\ref{eq:powerlawD}) and for large
masses $m^2\gg 1$, one expects a power-law dependence
$\alpha\propto m^{-7}$. Indeed, for large $m^2$ we get from
(\ref{eq:hamiltf}) $\epsilon\sim m^2 q^2 $, so that, at fixed
$\epsilon$, $q$ scales like $1/m$. Therefore the strength $g$ of
the quartic potential (and hence $x=g\epsilon$) scales like
$m^{4}$. Considering that $[D]=t^{-1}$ and the characteristic
oscillation time (\ref{eq:Tmicro}) scales like $1/m$, an
additional factor $m$ is obtained, yielding
\begin{equation}\label{eq:alpham}
    \alpha\propto m^{-7}, \qquad \mbox{for} \quad m^2\gtrsim 15 \ .
\end{equation}
As shown in Figure \ref{fig:massafit}, this prediction is well
confirmed by our numerical data, provided $m^2\gtrsim 15$. For
$0.1\lesssim m^2\lesssim 15$ we numerically found $\alpha\propto
m^{-3.3}$, for which we offer no explanation. Moreover, for $m\to
0$ one expects a qualitatively different situation, since
$\omega_{\mathrm{MAX}}/\omega_{\mathrm{min}}=\sqrt{4+m^2}/m\to\infty$,
such a ratio representing the available primary resonances between
normal modes \cite{Lichtenberg}. Actually, we observed the
formation of metastable states for $m^2 \lesssim 0.05$ (not shown
in Figure \ref{fig:massafit}): as a consequence, the correlation
function showed marked oscillations of definite frequency,
hindering a consistent definition of a diffusion coefficient.

\section{\label{sec:concl} Conclusions}
The analytic (power-law) divergence (\ref{eq:tau2x}) of the
intermediate timescales describing the Brownian motion on the
constant energy surface suggests that a perturbative approach,
based on a Liouville-Fokker-Planck equation
\cite{Prigogine60} should apply, at least \emph{locally}, and that
an eventually non-analytic divergence of the relaxation time from
far to equilibrium should be ascribed to a nontrivial structure of
the phase space at larger scales. The dependence of the
intermediate timescale on the mass and the presence of long-lived
metastable states indicates that the problem is very involved also
at the (supposedly simpler) level of the local dynamics in phase
space. In this context, one might speculate that the relaxation
from states that are not too far to equilibrium depends on the
``mesoscopic" features of phase space. (The expression
``mesoscopic" is not used here in its most familiar meaning,
related to the interplay of classical and quantum effects, but
rather in the sense of intermediate between ``microscopic" and
``macroscopic," i.e.\ pertaining to the total system.) The
approach we propose enables one to extract sensible information
from the dynamics at intermediate timescales (from close to
equilibrium) that are usually less studied than the longer ones,
describing the relaxation from far to equilibrium.  This has
obvious positive spinoffs from the perspective of numerical
investigations, as it requires less computing time. From a more
conceptual viewpoint, we have proved the existence of a diffusive
process (without introducing any randomization ``by hand"), that
constitutes a building block for the global equilibration process.

\appendix*
\section{}
\label{sec-appA}
Let us show that, in order to compute the average of any function
of the spectral entropy $S$, the microcanonical and the canonical
ensemble are equivalent. This conclusion is a consequence of the
following theorem:

\emph{In an ensemble of $N$ noninteracting integrable classical
systems, where the energies $E_i>0$ are homogeneous functions of
the action variables $I_i$, the average of functions of the kind
$F(E_i/E)$, where $E=\sum E_i$, computed according to the
Boltzmann distribution, is equal to that computed according to the
microcanonical distribution. Moreover, this average is independent
of the temperature $\beta$ of the canonical ensemble and the
energy $\cE$ of the microcanonical ensemble.}

We prove this theorem by calculating the canonical average of a
quantity $F$ (the angle variables are trivially integrated over
since they do not contribute to the energies)
\barr
\left< F\right>_c&=&\frac{1}{Z_c}\int
d^N\;Ie^{-\beta E}F\left(\frac{E_i}{E}\right)\nonumber\\
&=&\frac{1}{Z_c}\int d^NE\;J(E_i)e^{-\beta
E}F\left(\frac{E_i}{E}\right),
\earr
where
\begin{equation}\label{eq:Zc}
 Z_c=\int d^N I\; e^{-\beta E}
\end{equation}
is the canonical partition function and $J=\|\partial I_j/\partial
E_i\|$ the Jacobian. Due to our hypotheses, this is a homogeneous
function of the $E_i$'s. We obtain
\barr
\left< F\right>_c&=&\frac{1}{Z_c}\int d^NE\int_0^\infty d\xi
\nonumber\\
& &\times \delta\left(\xi-\sum_i E_i\right)e^{-\beta E}J(E_i)F
\left(\frac{E_i}{E}\right)\nonumber\\
&=&\frac{1}{Z_c}\int_0^\infty d\xi\; e^{-\beta\xi}\xi^{N-1}
\nonumber\\
& &\times \int d^Nx\;J(\xi x_i)\delta\left(1-\sum_i
x_i\right)F\left(x_i\right)\nonumber\\
&=&\frac{1}{Z_c}\int_0^\infty d\xi\;
\xi^{N-1+N\alpha}e^{-\beta\xi}
\nonumber\\
& &\times \int d^NxJ(x_i) \delta\left(1-\sum_i
x_i\right)F\left(x_i\right)\nonumber\\
&=&\frac{1}{Z_c}\beta^{-N(1+\alpha)}\Gamma(N(1+\alpha))\nonumber\\
& &\times \int d^NxJ(x_i) \delta\left(1-\sum_i
x_i\right)F\left(x_i\right),
\earr
where $\Gamma$ is the Gamma function, we defined $x_i=E_i/\sum_j
E_j$ and used the homogeneity of the Jacobian to write
\beq
J\left(\xi x_i\right)=\xi^{N\alpha}J(x_i),
\eeq
which defines the quantity $\alpha$. By repeating the same
derivation with $F=1$, one readily shows that
\beq
\label{eq:constZZ}
\frac{Z_c}{\beta^{-N(1+\alpha)}}=
\Gamma(N(1+\alpha)) \; \frac{Z_m}{\cE^{N-1+N\alpha}} ,
\eeq
where the $Z_m$ is the microcanonical partition function
\begin{equation}\label{eq:Zm}
 Z_m=\int d^N I\; \delta(\cE-E).
\end{equation}
Incidentally, notice that both sides of (\ref{eq:constZZ}) are
independent of $\beta$ and $\cE$. In conclusion
\barr
\left<F\right>_c&=&\frac{\cE^{N-1+N\alpha}}{Z_m}\int
d^Nx\;J(x_i)\delta\left(1-\sum_i x_i\right) F\left(x_i\right)
\nonumber\\
&=&\frac{1}{Z_m}\int d^N I\; \delta\left(\cE-E\right)
F\left(\frac{E_i}{E}\right) = \left<F\right>_m.
\label{eq:Zmc}
\earr
This proves our assertion. Equations
(\ref{eq:meanS})-(\ref{eq:microcandeltas2}) are readily obtained
by explicitly calculating the (simpler) canonical average of $F=S$
and $F=S^2$.

As a corollary, one sees that if the $E_i$s are random variables
distributed with $e^{-E_i}$, the variables $x_i=E_i/\sum_j E_j$
are distributed like $\delta(1-\sum_j x_j)$. This provides a fast
numerical recipe to generate microcanonically distributed
variables.

\end{document}